\documentclass[
    ,final            
  ]
  {aipproc}

\layoutstyle{6x9}

\begin{document}

\title{Observational evidence for supermassive black hole binaries}

\author{Stefanie Komossa}{
  address={Max-Planck-Institut f\"ur extraterrestrische Physik, Giessenbachstr. 1, 85748 Garching,
           Germany; skomossa@xray.mpe.mpg.de}
}

\begin{abstract}
Coalescing massive black hole binaries are powerful emitters of gravitational
waves, in the {\sl LISA} sensitivity range for masses $M_{\rm BH} \approx 10^{4-7}$\,M$_{\odot}$.
According to 
hierarchical galaxy merger models,  
binary black holes should form frequently, and should be common in the cores
of galaxies. 
The presence of massive black hole binaries has been
invoked to explain a number of class properties of different
types of galaxies, and in triggering
various forms of activity.
The search for such binary black holes is therefore of great interest for key topics
in astrophysics ranging from galaxy formation to activity in galaxies.

A number of phenomena were attributed to the presence of supermassive binary black holes,
including
X-shaped radio galaxies and double-double radio galaxies,
helical radio-jets, periodicities in the lightcurves of blazars,
(double-horned emission-line profiles), binary galaxies with radio-jet cores,
binary quasars, and the X-ray active binary black hole at the center 
of the galaxy NGC 6240.
Here, I review the observational evidence for the presence
of supermassive binary black holes in galaxies, and
the scenarios which have been
discussed to explain these observations.
\end{abstract}

\maketitle

\vspace*{-13.1cm}
{\footnotesize{
\begin{verbatim}
review, to appear in: `The Astrophysics of Gravitational Wave Sources',
J. Centrella (ed), AIP, in press   (2003) 
\end{verbatim}
}}
\vspace*{10.1cm}



\section{Introduction}

The major route of forming binary black holes is via mergers of galaxies.
If both merging galaxies harbor supermassive black holes at their centers
(e.g., Richstone 2002), these two black holes will
finally sink to the center of the potential and will eventually merge.
Hierarchical merger models of galaxy formation predict that
binary black holes should be common in galaxies (e.g., Haehnelt \& Kauffman 2002,
Volonteri et al. 2003).  The merging process will be enhanced in clusters
and groups of galaxies.
Under some circumstances, black holes may already form as binaries
at the centers of the first galaxies (Bromm \& Loeb 2003).

Mergers of massive binary black holes produce
strong gravitational wave signals
(e.g., Thorne \& Braginsky 1976, Centrella 2003, Baker 2003)
which will be detectable
for the first time with the future space-borne gravitational wave
interferometer mission {\sl LISA} (e.g., Bender et al. 1998, Danzmann 1996, 
Haehnelt 1994) for masses in the 
range $\sim$10$^{4-7}\,M_{\odot}$.

Supermassive binary black holes (BBHs) are expected to 
be of wide astrophysical relevance.
Their detection and number estimates provide important constraints
on models for galaxy formation and evolution.
Their detection allows to study one possible route of BH growth, that
by merging of BHs with each other, the frequency of these events,
and their relevance for BH growth.
BBHs were suggested to play a role in increasing AGN activity
(e.g, Gaskell 1985, Gould \& Rix 2000), 
in triggering starburst activity (Taniguchi \& Wada 1996),
and in the formation of molecular tori (Zier \& Bierman 2001, 2002)
which are believed to be an important ingredient in
unified models of AGN.
The possible relevance of BBHs in explaining different classes of radio-loud
AGN was variously addressed (e.g., Basu et al. 1993, Wilson \& Colbert 1995, Villata \& Raiteri 1999,
Britzen et al. 2001), 
and it was suggested that the misalignments between the direction
of radio jets and disks in active galaxies (AGN) are due to past black hole merges in
these galaxies (Merritt 2002, 2003). 
Once {\sl LISA} results become available, the gravitational waves from coalescing binary
black holes will be used to infer merger rates and the merger history of BHs/galaxies
(e.g., Haehnelt 1994,
Menou et al. 2001, Hughes 2002, Menou 2003),
and possibly black hole masses (Hughes 2002, Menou 2003). In the future,
gravitational waves of coalescing BHs may be used as cosmological standard candles,
if the optical counterparts of these systems can be identified (Holz \& Hughes 2003).

Given the importance of BBHs, the wide role they may play in explaining
observations of classes of activity, and given the expectations that they
should be common, based on galaxy formation scenarios, one key
question is: what is the actual
observational evidence that BBHs do exist ? This is the topic of this review.
An overview is given of the different types of observations which
point to the presence of BBHs, and of the scenarios invoked to explain these
observations. Remaining uncertainties in both, observations
and models are discussed. 
Due to space limitations, referencing will be incomplete. My apologies
in advance.

\subsection{Binary black hole formation and evolution}

BBHs will be formed in the course of merging
of two galaxies (e.g., Begelman et al. 1980, Valtaoja et al. 89,
Milosavljevic \& Merritt 2001, Yu 2002, and references therein).
The merging of the two black holes basically proceeds in three stages
(e.g., Fig. 1 of Begelman et al. 1980).
In a first stage, the cores of the
merging galaxies reach closer towards each other
by dynamical friction (e.g. Valtaoja et al. 1989).
The third stage is the actual merging of the two black holes by
emission of gravitational waves (GWs). Which processes are effective
in the intermediate second stage and how efficiently they operate,
i.e., how quickly the separation between the BHs shrinks to a
distance where GW emission becomes significant,
is still a subject of intense theoretical study (see Merritt 2003 for a review).

A number of processes which could lead to a hardening of the black hole
binary
were discussed, including
stellar slingshot effects and re-filling of the loss cone (Saslaw et al. 1974,
Quinlan \& Hernquist 1997, Milosavljevich \& Merritt 2001, Zier \& Biermann 2001,
and references therein), black hole wandering (e.g. Merritt 2001, Hemsendorf et al. 2002,
Chatterjee et al. 2003), interaction with surrounding gas and the accretion disk
(e.g. Ivanov et al. 1999, Gould \& Rix 2000, Haehnelt \& Kauffmann 2002,
Armitage \& Natarajan 2002),
and/or the Kozai mechanism (Blaes et al. 2002).
Theoretical calculations show, that under some circumstances none of
these mechanisms may operate in sufficient strength, with the consequence that
the BBH is expected to stall at separations of 
0.01-1 pc (e.g., Milosavljevich \& Merritt 2001). 

There is
circumstantial
evidence that most BBHs do actually merge in less than a Hubble time.
Haehnelt \& Kauffman (2002) argued,
if the binaries lasted too long, some of them would be ejected
from the nucleus in the course of a three-body
interaction with a third black hole, once a new merger occurs. That
would lead to the prediction of the existence of galaxies without BHs at their center,
as opposed to observations that most galaxies do harbor SMBHs.

Theoretical issues of BBH formation and evolution are covered in much greater detail 
by Haehnelt, Volonteri et al., and Milosavljevich \& Merritt in these
proceedings, and in the review by Merritt (2003).

Most of the BBH evidence described below is based on observations of 
{\em active} black holes, i.e. phenomena like
presence of
radio-jets, presence of hard and luminous X-ray emission, and optical emission-line ratios
typical of Seyfert galaxies.

\subsection{BBHs as explanations for different classes of astrophysical objects
  or specific components of the AGN core}

It is generally expected, that merging between galaxies may trigger
all kinds of activity in the center of the merger remnant (e.g., Gaskell 1985),
for instance by driving gas to the central region (Barnes \& Hernquist 1996)
which then triggers starburst and AGN activity.

In particular, the presence of a {\em binary} black hole 
was explicitly invoked to explain different
facets of activity in galaxies:
Taniguchi \& Wada (1996) proposed that the BBH which was formed in the course of a major merger
triggers starburst activity near the nucleus, while BBHs in minor mergers lead to
formation of hot-spot nuclei.   

According to Zier \& Bierman (2001, 2002), the  influence of the BBH
on the surrounding stellar distribution will lead to a torus-like structure.
Winds of these stars would then form the tori of molecular gas,
thought to be ubiquitous in AGN and thought to play an important role
in unification scenarios of AGN (Antonucci et al. 1993). 

The possible relevance of BBHs in explaining different classes  
of radio-loud
AGN was addressed repeatedly 
(e.g., Basu et al. 1993, Wilson \& Colbert 1995, Villata \& Raiteri 1999,
Valtonen \& Hein\"am\"aki 2000, Britzen et al. 2001, Merritt 2002).
Villata \& Raiteri (1999) speculated that all blazars "owe their origin to the presence
of binary black holes", and that differences between different classes of AGN are due to an
evolutionary sequence:  BL Lacs and FRI radio galaxies would represent the advanced stages
with close binary pairs and low mass accretion, while FRII galaxies would harbor the wide pairs
with long orbital periods (observationally less easily recognizable). 

Wilson \& Colbert (1995) suggested that the difference between radio-loud and radio-quiet AGN
arises because the former, in elliptical galaxies,
posses rapidly spinning BHs, spun  up
(not by accretion but) as a result of the coalescence of the two
original BHs, while the latter, residing in spiral galaxies, have non-spinning BHs.
The jets are considered to be powered by the spin of the BHs in the radio-loud galaxies.

Merritt (2002, 2003) pointed out that the spin re-orientation of the primary
black hole caused by repeated mergers would explain the fact that the orientations
of radio jets of AGN are almost random with respect to the plane of the stellar disk.   

Finally, different types of jet structures in individual galaxies were
interpreted as observational evidence for the presence of BBHs. 
This topic is addressed in more
detail in the next sections.

\section{Observational evidence for binary black holes: \newline
              spatially unresolved systems}

\subsection{BBH merger remnants}

\subsubsection{X-shaped radio galaxies}

Some radio galaxies show jets with very peculiar
morphology; abrupt changes in jet direction,
forming X-shaped patterns (e.g., Fig. 2 of Leahy \& Williams 1984,
Fig. 1 of Parma et al. 1985, Leahy \& Parma 1992, Capetti et al. 2002, Wang et al. 2003).
The changes are more abrupt than in S-shaped radio galaxies. While the latter shapes
are usually explained in terms of precession effects (e.g. Ekers et al. 1978),
Parma et al. (1985) noted that the cross-shaped morphology is likely linked to
an abrupt change in jet ejection axis, or is consistent with a precession phenomenon
for a specific viewing geometry.

About 15 X-shaped or `winged' radio sources are known, most
of them associated with low-luminosity FRII sources.
Except one (Wang et al. 2003), none of the X-shaped radio galaxies
shows optical quasar activity in form of {\em broad} emission lines at their center.
The host galaxies mostly exhibit high ellipticities (Capetti et al. 2002).
A number of them have companion galaxies (Tab. 1 of Merritt \& Ekers 2002),
and the host galaxy of one (3C293)
shows clear signs of interaction (Evans et al. 1999).

Several different models to explain the X-shaped patterns
were addressed in the literature, which either link the characteristic
X-shape to a re-orientation of the jet axis, or to effects of backflow
from the active lobes into the wings
(see, e.g., Sect. 5,6 of Dennett-Thorpe et al.
2002 for a summary){\footnote{Leahy \& Williams
(1984) studied a backflow
model to explain the X-shaped radio structures 
(their Fig. 6; see Capetti et al. 2002 for another variant
of these type of models). In their model
of a light jet, post-jet-shock material is re-accelerated towards
the center, causing backflows which, back at the source,
expand laterally.
In favor of backflow type of models,
as opposed to jet-reorientation models, Capetti et al. (2002) noted that
a number of the winged radio galaxies show wings preferentially aligned
with the minor axis of the host galaxies.
On the other hand, Dennett-Thorpe et al. (2002) argued that backflow
models are untenable for sources like 3C223.1 and 3C403 which show
wings that are longer than the active lobes.
Other types of models, argued to be unlikely by Merrit \& Ekers (2002),
 involve accretion of dwarf galaxies  
and warping instabilities
of the accretion disk in order to explain a re-orientation of the jet axis.
}}.

According to Merritt \& Ekers (2002) and Zier \& Biermann (2002),
the X-shaped patterns reflect changes in the orientation
of the black hole's spin axis, caused by the merger with a second
SMBH. This is also one of the two explanations preferred by Dennett-Thorpe et al. (2002). 
Merritt \& Ekers favored minor mergers and showed that these can 
significantly change the spin axis of the primary black hole.

\subsubsection{Double-double radio galaxies with interrupted jet activity}

Secondly, so called double-double radio galaxies (Schoenmakers et al. 2000)
were suggested to be
remnants of merged BBHs (Liu et al. 2003). These are sources which
exhibit pairs of symmetric double-lobed radio-structures, aligned along the same axis.
Inner and outer radio lobes have a common center (Fig. 1-3 of Schoenmakers et al. 2000)
and there is a lack of radio emission between inner and outer lobes.
The most likely origin of these structures is an interruption and re-starting
of the jet formation.  The interruption time scale is about 1 Myr. 

Several
ideas were proposed to explain this phenomenon (see Sect. 5 of Schoenmakers
et al. 2000). Liu et al. (2003) favor the presence of binary
black holes which have already merged in these systems.
According to Liu et al. the inward spiralling secondary black hole temporarily leads
to the removal of the inner parts of the accretion disk around the primary black hole, thus to
an interruption of jet formation. Jet activity restarts
after the outer parts of the accretion disk refill the
inner parts of the disk.

\vskip0.4cm

These two important classes of candidates for BBH merger remnants --  
the X-shaped radio galaxies and the
double-double radio galaxies -- will certainly receive intense 
observational interest in the next few years, in order to address
questions like: are there any  double-double radio galaxies with {\em changes} in the
jet direction (Liu et al. assumed the merging BHs to be coplanar
but this does not always have to be the case) ? 
Do we see X-shaped radio galaxies with an interruption of jet-activity
between the core and the wings (if the wings reflect an earlier period
of activity they would no longer be powered today, which could be confirmed
by very high-resolution radio imaging) ?

\subsection{Helically distorted radio jets}

A third phenomenon exhibited by radio jets 
which was suggested to be linked to the presence
of binary black holes (Begelman et al. 1980)
is the presence of (semi-periodic) deviations of the jet directions
from a straight line. 
Helical distortions and bendings of jets have been seen 
in a number of sources (including 3C273, 3C449, BL Lac, Mrk 501, 4C73.18
and PKS0420-014;
see e.g., Fig. 1 of Roos et al. 1993, 
Fig. 3 of Tateyama et al. 1998), interpreted as manifestation of BBHs in these systems. 

While there is general agreement that the wiggles in radio jets are most
plausibly caused by the presence of BBHs, different mechanisms
to explain the observations have been favored. These either link the observations to
{\em orbital motion} of the jet-emitting black hole  (e.g., Kaastra \& Roos 1992,
Roos et al. 1993,
Hardee et al. 1994),
or to {\em precession effects}, either precession of the accretion
disk around the jet-emitting black hole under gravitational torque (on shorter time scales;
e.g. Katz 1997, Romero et al. 2000),
or to geodetic precession (acting on longer time scales; e.g. Begelman et al. 1980).
Correspondingly, black hole mass estimates in these systems are still
subject to uncertainties by a factor $\sim$10-1000. 

\subsection{Semi-periodic signals in lightcurves}

\subsubsection{OJ 287}

Another periodic phenomenon very often attributed to the presence of
BBHs is (semi)periodic changes in lightcurves.
The best-studied candidate for harboring a BBH, inferred from the characteristics
of its optical lightcurve, is the BL Lac Object OJ 287, which exhibits
optical variability with quite a strict period of 11.86 years
(Silanp\"a\"a et al. 1988, 1996, Valtaoja et al. 2000, Pursimo et al. 2000,
and references therein).
Optical observations of this source can be followed back to 1890
(e.g., Fig. 1 of Pursimo et al. 2000).
How can the BBH model explain periodic optical variability ?
Basically, two different classes of models were discussed:
(i) accretion(disk)-related variations in the luminosity (e.g., Silanp\"a\"a et al. 1988,
Lehto \& Valtonen 1996), or (ii) jet-related
variability due to Doppler-boosting of varying strength
(Katz 1997, Villata et al. 1998).
Variants of
them have also been invoked to explain apparent periods in
the data of other BL Lac objects.
(i) According to the original idea of Silanp\"a\"a et al. (1988), the
tidal perturbation, when the secondary
approaches closest to the accretion disk of the
primary leads to increased
accretion activity, thus a peak in the optical lightcurve.
(ii) Katz et al. (1997) studied a model in which precession of the accretion disk,
driven by the gravitational torque of a companion mass (a second black hole),
causes the jet to sweep periodically close to or across our line of sight. This then leads to
a modulation of the observed intensity of light due to Doppler-boosting. 

Combining radio and optical observations of OJ 287, Valtaoja et al. (2000)
recently favored a scenario in which the first optical peak is
due to a thermal flare, when the secondary BH plunges
into the accretion disk of the primary.
Following the major optical flare, a second optical flare is observed about
a year after the first peak, and (only) that second peak is accompanied by enhanced radio emission
(Valtaoja et al. 2000). This second peak
is traced back to the tidal perturbation, excerted by the secondary black hole
upon closest approach to the accretion disk of the
primary which leads to increased
accretion activity.
The observed $\sim$12 yr period ($\sim$9 yr in the rest frame)
then corresponds to the orbital period
of the BBH.

A possible alternative to the BBH models, a well-defined duty-cycle model of accretion,
is briefly mentioned by Valtaoja et al. (2000), but is considered as an unlikely
explanation for the case of OJ 287, due to its  quite strict periodicity.

The next optical maximum of OJ 287 is expected in March 2006. No doubt, this BL Lac will
then be the target of intense multi-wavelength monitoring campaigns.

\subsubsection{Other cases}

Optical variability with indications for periodicity ($N$ = a few, where $N$ is the number
of periodic oscillations) and
periods on long (order 10-20 yrs) and intermediate
(order 20 days - 1 yr) scales have been observed in
other blazars (see Sect. 1 of Fan et al. 1998 and Tab. 1 and 2 of Xie 2003 for summaries).
Among them a 336 day period of minima in the optical lightcurve of
PKS1510-089 (Xie et al. 2002), a 14 yr period in optical data of BL Lac (Fan et al. 1998),
a 5.7 yr period of AO 0235+16 at radio frequencies
(Raiteri et al. 2001), and a 23-26 day period 
of Mrk 501 at TeV energies (Hayashida et al. 1998).
No other data set brackets as long a time
interval as OJ 287, though, and periodicities are generally
less conspicuous and less persistent
as in OJ 287{\footnote{For instance, the lightcurves of BL Lac also indicate
periods of 0.6 yr, 0.88 yr, 7.8 yr and 34 yr, and it is unclear whether all of them are real,
and what causes them.}}.

In some, but not all cases, the variability in the lightcurves was traced back to the presence of
a BBH, but different types of models were involved, which either relate the variability
to real changes in the luminosity, or apparent changes due to Doppler-boosting effects in
those lightcurves which show periodic brightness peaks.

Rieger \& Mannheim (2000, see DePaolis et al. 2002 for a generalization
of their approach to other orbits) showed that the
periodic variability of the BL Lac object Mrk 501 at TeV energies with a period of
about 23 days and $N$=6 is possibly related 
to the presence of a BBH. According to their model,
the observed flux modulation arises from a varying 
Doppler-factor due to slight variations
in inclination angle of (a moving blob in) the jet, 
caused by the orbital
motion of the less massive, jet-emitting black hole.
The presence of
a BBH in Mrk 501 was also discussed with respect to the complex radio
jet morphology of this source
(Conway \& Wrobel 1985, Villata \& Raiteri 1999),
and in relation to peculiarities in its spectral
energy distribution (Fig. 2 of Villata \& Raiteri 1999).

\subsection{Doule-peaked BLR line profiles}

If binary black holes exist in active galaxies,
it can plausibly be expected that
several are at nuclear separations such that their orbital motion causes observable
effects on the profiles of the broad emission lines (e.g. Stockton \& Farnham 1991,
Gaskell 1996).

A number of AGN were observed which show  
double-horned emission-line profiles (Arp 102B and 3C390.3
are prominent 
examples). 
These line profiles were interpreted as evidence for two 
physically distinct broad line regions (BLRs)
of two black holes at the centers of the two galaxies.  

However, in those cases examined closely, so far, the BBH interpretation was disfavored,
mostly because the predicted temporal variations in the line profiles, as the two
BHs orbit each other, were not detected in optical spectroscopic monitoring programs 
(Halpern \& Filippenko 1988, Eracleous et al. 1997). In one
case, the apparent double-horned profile of the Balmer lines turned out to be
an artifact of other spectral peculiarities (Halpern \& Eracleous 2000).

Yu (2002) pointed out that orbital timescales of binary BHs with BLRs in AGN
may generally be expected to be as large as 100-1000 yr, in which case no
detectable variability in the red and blue peak of the emission-line profile is expected
on the timescale of years.
A further complication in the search for BBHs via characteristic line profiles
is the existence of a number of other mechanisms which are 
also known to produce double-peaked line profiles, 
like bipolar outflows and accretion disks, not related to BBHs.

\subsection{Galaxies which lack central cusps}
Essentially all evidence presented for the presence of BBHs is linked
to some activity of at least one of the two black holes
which form the pair (like radio jets, AGN-like emission-lines, etc.).

There is one exception, which is some rather indirect hint
for the presence of binary black holes.
Lauer et al. (2002) used {\sl HST} to identify several early-type galaxies 
with inward-decreasing surface-brightness profiles.
The presence of galaxies with such central `holes' rather than cusps is
expected in some models of BBH evolution (see Sect. 6 of Merritt 2003
for a summary), reflecting the ejection of stars from the core 
in the course of  
the hardening of the black hole binary.

\section{Observational evidence for binary black holes: spatially resolved systems}

If a large number of BBHs exist in very close orbits, we also expect some pairs
of larger separations, directly resolvable as individual SMBHs in nearby
galaxies. Below, I discuss the few available observations which fall
in this category.

\subsection{X-ray active black hole pair: NGC 6240}

Recent observations of the {\sl Chandra} X-ray observatory let to the
discovery of a pair of active
black holes{\footnote{Actually, what is observed is the characteristic
signs of AGN activity. The step that this is
equivalent to SMBH detection is based on the widespread belief,
that AGN are indeed powered by SMBHs. This statement basically holds for
all other BBH evidence: it is always the activity associated
with the black hole(s) which is detected.}} in the center of the (ultra)luminous
infrared galaxy NGC 6240 (Komossa et al. 2003).
The BBH in NGC 6240 is special in that it is presently the {\em only} BH pair
I am aware of at
the center of one galaxy, and spatially resolved such that both (active) BHs can
be separately identified.

The galaxy NGC\,6240 belongs to the class of (ultra)luminous infrared galaxies (ULIRGs)
which are characterized by an IR luminosity exceeding $\sim$10$^{12}$\,L$_{\odot}$
(see Sanders \& Mirabel 1996 for a review).
NGC\,6240 is one the nearest (U)LIRGs and is considered a key representative
of its class. NGC\,6240 is the result of the merger of two galaxies, expected
to form an elliptical galaxy in the future. It harbors two optical nuclei
(Fried \& Schulz 1983). Their nature remained unclear. In particular, no optical
signs of AGN activity showed up in ground-based optical spectra.
An active search for obscured AGN activity in NGC\,6240 was carried
out during the last two decades (see Sect. 1 of Komossa et al. 2003 for a summary)
which let to the detection of absorbed, 
intrinsically luminous X-ray emission (Schulz et al. 1998)
extending up to $\sim$100 keV (Vignati et al. 1999) from the
direction of NGC\,6240.
Spatial resolution was insufficient to locate the source of the
hard X-ray emission, though.    

Employing the superb spatial imaging spectroscopy
capabilities of the X-ray observatory {\sl Chandra}, it was found that {\em both} nuclei
of NGC\,6240 are active (Fig. 1), i.e. harbor accreting supermassive black holes
(Komossa et al. 2003).  The southern and northern nucleus show very similar
X-ray spectra which are flat, heavily absorbed, and superposed on
each is the presence of a strong neutral (or low-ionization) iron line. 
These kind of spectra have only been observed in AGN.  
 
The projected separation of the X-ray cores is 1.5 arcsec, corresponding to a
physical separation of 1.4 kpc.
Over the course of the next few hundred million
years, the two black holes in NGC6240 are expected to merger with each other.

\begin{figure}
  \includegraphics[height=.36\textheight]{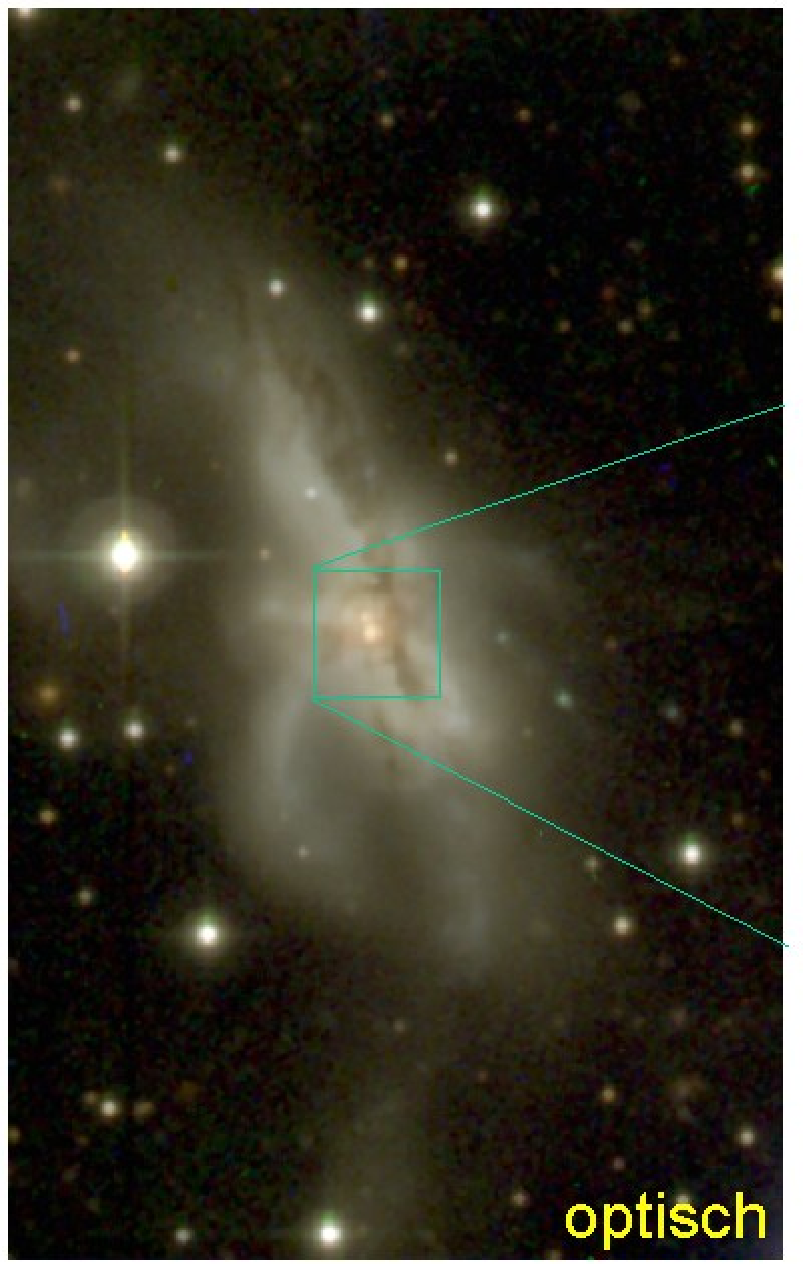}
 \vspace{-6.0cm}\hspace{1.6cm}
  \includegraphics[height=.35\textheight]{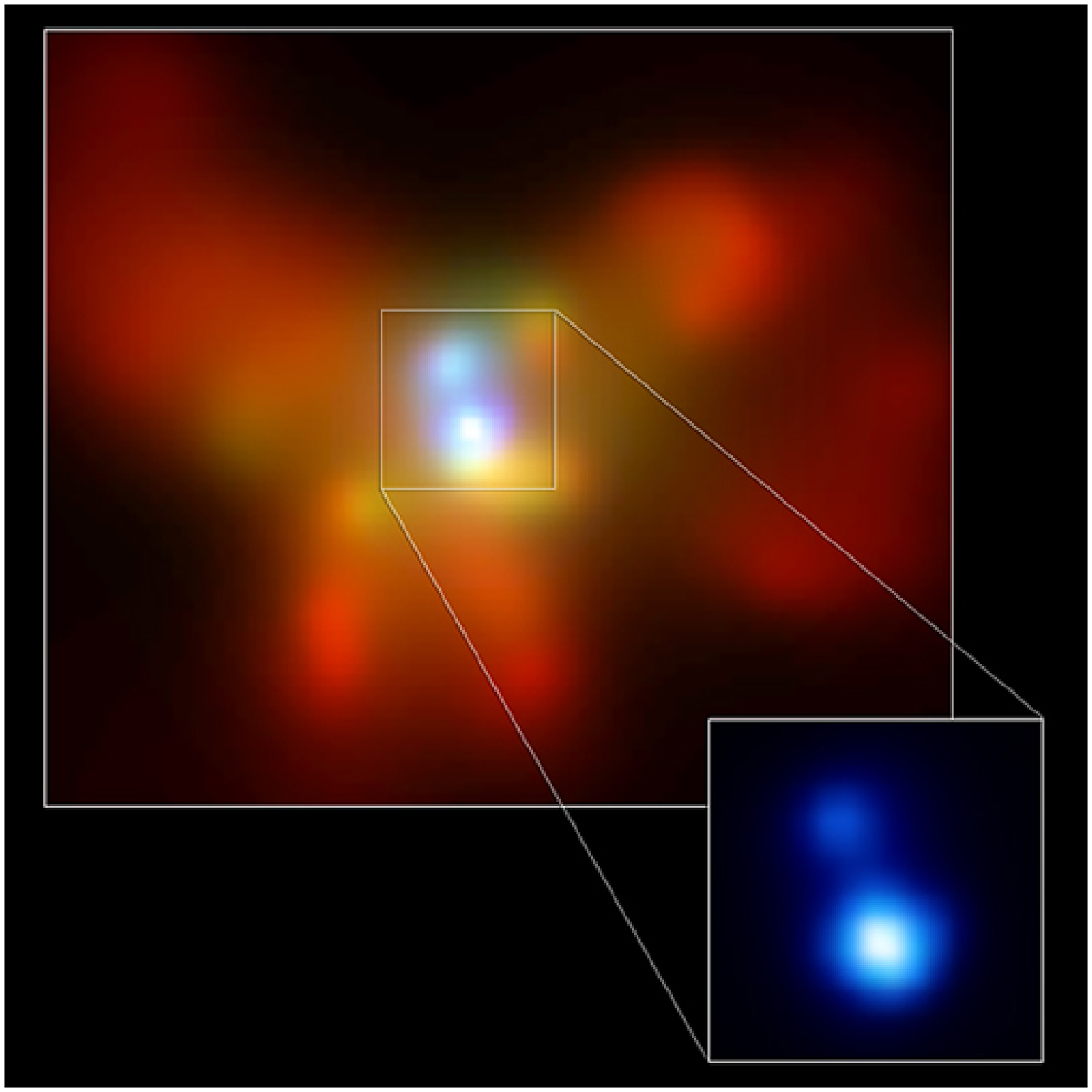} 

  \caption{Optical image of NGC\,6240 (right; Keel 1990) and X-ray image of 
the galaxy's center (left; Komossa et al. 2003). 
The X-ray image is color coded, reproduced here in black and white. 
The extended emission is soft, while the compact emission is harder. 
The inset zooms onto
the two hard active nuclei of this galaxy.}
\end{figure}

\subsection{Binary quasars}

A small fraction of (high-redshift) quasars were observed to be accompanied by
a second nearby quasar (image) with the same redshift.
These may either be gravitational lenses, true quasar pairs, or chance alignments.
The majority of them are in fact the result of gravitational lensing.

Several pairs with separations 3-10$^{\prime\prime}$
were argued to be real pairs, i.e., physically distinct
binary quasars (see, e.g., Tab. 1 of Mortlock et al. 1999 and of Kochanek et al. 1999,
Sect. 2.5.8 of Schneider et al. 1992).
The distinction between a real pair and a lensed quasar is mostly
based on: (i) whether or not a lens is detected
and (ii)  whether or not the quasar spectra differ significantly, and/or whether or not
there are other significant differences, like when one image is radio-loud,
the other radio quiet.
Lack of a detectable lens, and very different quasar spectra/properties strongly argue for
real pairs. There is a growing number of quasar images identified
as
true pairs or as excellent candidates for true pairs,
among them
Q1343.4+2640 (with substantial spectral differences; Crampton et al. 1988),
LBQS0103-2753 (the smallest-separation pair known,
with very different optical spectra; Junkkarinen et al. 2001),
LBQS0015+0239 (interpreted as probable binary system  due to lack of an optically bright lens
which then makes it the highest-redshift binary known, at $z$=2.45; Impey et al. 2002),
LBQS1429-0053 (similar optical spectra but lack of a lens candidate; Faure et al. 2003),
UM\,425 (similar optical/UV
spectra but different amounts of absorption in X-rays and lack of a bright lens;
Mathur \& Williams 2003, Aldcroft \& Green 2003),
and Q2345+007 (Green et al. 2002).
Q2345+007 at $z$=2.15 is an interesting case: Despite essentially
identical optical spectra, no lens could be found in the optical and X-ray band (Green et al. 2002).
The X-ray spectra of the quasars do differ, and Green et al. favor an interpretation
in terms of  a real quasar pair.

The smallest-separation known binary quasar is LBQS0103-2753 at redshift $z$=0.848.
Its projected
separation of 0.3$^{\prime\prime}$ corresponds
to $\sim$ 2.5 kpc.
It remains uncertain, whether the projected separation reflects the true separation.
For instance, the emission-line shifts in both quasar spectra ($v$=3900 km/s)
indicate a much larger separation, but emission lines are not always good indicators
of systemic redshift (Junkkarinen et al. 2001). Junkkarinen et al. point out that if the
systemic redshift difference is much smaller than indicated by the emission lines,
LBQS0103-2753 would most likely be a galaxy merger with two quasar cores.

It is interesting to note that there appears to be a lack of small pair separations.
Apart from LBQS0103-2753 other pairs have separations $>$ 2$^{\prime\prime}$, and typically 3-10$^{\prime\prime}$.
These projected separations
convert to
distances of $\sim$10-80 kpc, given the redshifts of the quasars (see Fig. 7 of Mortlock et al. 1999
for an attempt to derive physical separations by a random deprojection method).

In any case, a chance projection of the quasar pairs on average is very unlikely, because
the pairs are in excess of what is expected from chance projections of single, unrelated sources.
Open questions then are: are the host galaxies of the quasars interacting, are they
bound to each other and in the process of merging ?
At low redshifts, this question can be addressed by searching for morphological
and kinematical distortions in the host galaxies. However, most binary quasars
are at large redshifts ($z \approx 1-2$), and sometimes the host galaxies
are not even detected{\footnote{Kochanek et al. (1999b) used HST to find that the
host galaxies of the binary quasar MGC 2214+3550 are undisturbed,
implying their physical separation is greater than their projected separation of $\sim$20 kpc}}.

If the quasar cores reside in galaxies which are already interacting with each other,
then the quasar activity might have plausibly been triggered in the
course of the interacting/merging process (e.g., Kochanek et al. 1999).

Assuming the observed binary quasars are bound pairs,  Mortlock et al. (1999)
estimated that dynamical friction will drive them closer together on a timescale
comparable to a Hubble timescale. Since this is longer than the typical activity time scale
of quasars, these authors pointed out that closer pairs may already be beyond the phase
of quasar activity (no fuel left in the center) and therefore not easily detectable.

\subsection{Pairs of galaxies which both harbor radio-jet sources: 3C75}

In the course of a VLA radio survey of Abell clusters,
the very unusual morphology of the radio source 3C75
close to  the center of the cluster of galaxies Abell 400 was discovered.
It showed {\em two} pairs of radio jets (Fig. 1 of Owen et al. 1985)
and was first considered an apparently
single galaxy with two cores.

However, optical follow-up observations showed two elliptical galaxies which each
posses a radio-jet emitting core.  Such a configuration is quite unusual,
and an important question became: are these two galaxies interacting or
merging with each other ? Generally, this can be decided upon examination
of the isophotes and kinematics of the galaxies.
For instance, asymmetric isophote distortions and tidal  tails
indicate an ongoing (advanced) merger (e.g. Lauer 1988).
As regards 3C75, no distortions of the inner isophotes and no kinematic disturbances
were found (Balcells et al. 1995, Govoni et al. 2000).
However, the outer isophotes show an off-centering and twist
(Lauer et al. 1988, DeJuan et al. 1994, Balcells et al. 1995, Govoni et al. 2000). These
observations indicate, that the two galaxies are presently not bound to each other,
but do interact with each other (Balcells et al. 1985).
Balcells et al. cautioned, that dust may play a role as well
in explaining the non-concentric isophotes.
Independently, though, it was argued
that the galaxies likely are not spatially very distant from each other, because the
jets show similar bends, i.e. should have passed similar
regions of the ICM (Owen et al. 1985).

In any case the two elliptical galaxies are close to the cluster center, and are expected
to sink to the center of the potential and merge, ultimately.

\section{Concluding remarks, LISA rates, uncertainties, future observations}

In summary, various lines of evidence point to the existence of supermassive
binary black holes at the centers of galaxies.
According to BBH evolution models, the longest timescales in the evolution
of the binary up to coalescence  are likely those in which
the binary is closely bound ($\sim$ 0.01-10 pc; see, e.g., Fig. 1 of Begelman et al. 1980);
most BH pairs would therefore not be spatially resolved.
These pairs likely manifest their presence in imprinting periodic
phenomena on observed galaxy properties, like wiggling radio jets and quasi-periodic
variations in lightcurves (most prominently seen in OJ 287).

However, we do expect to see some wider pairs which can be spatially resolved.
It should be noted, though,
that these can only be easily recognized, {\em if both black holes are active}. Likely,
a number of them escape detection if the second BH is not in an active state at
the epoch of observation.
The pair with the smallest physical separation spatially resolved ($\sim$1.5 kpc), 
and the only one we are aware of at the center of a single galaxy, is the X-ray active pair
of black holes in NGC 6240.
There are wider pairs, in form of binary galaxies with active centers, like,
the two galaxies which make the radio source 3C75
or several binary quasars; those with
closest spatial separations may already be in an advanced stage of merging.

Some post-merger candidates for BBHs were proposed, especially the X-shaped radio
galaxies, interpreted as result of minor mergers between two black holes, leading to
a spin-flip in the resulting black hole and a corresponding change in jet direction.
Based on this scenario, Merritt \& Ekers (2002) predict a merger event rate 
for this type of galaxies of $\sim$1/year
detectable by gravitational wave interferometers. 

Within the framework of the BBH models,  uncertainties arise from questions like, 
which scenario is the correct one
(in particular, orbital motion versus precession effects) to 
describe the data. This, in particular, leads
to large uncertainties in predictions of the BH masses of the observed systems,
which range from 10$^6$ - 10$^{10}$ M$_\odot$.  The high BH mass range would 
be outside the {\sl LISA} sensitivity range. 

Future search for (active) binary BHs will likely concentrate on topics like
(i) an extended X-ray search for active binary black holes at the centers of 
(ultraluminous IR) galaxies using {\sl Chandra}, 
(ii) deep {\sl HST} imaging of the host galaxies of
quasar pairs at high redshift, to search for interaction-induced morphological
distortions, (iii) high-resolution radio imaging of X-shaped radio galaxies
to search for interruption of the radio emission between core and wings,
and (iv) multi-wavelength monitoring of blazars with (suspected)
periodical variability, in particular an intense coverage of the next expected
maximum of OJ 287.  

Once {\sl LISA} operates, it will provide us with valuable complementary information
on the rate and merger history of binary black holes. 
Ideally, if {\sl LISA} sources could be identified with optical counterparts,
gravitational wave signals and observations across the electromagnetic spectrum could be
combined to study BBH merger systems in detail.

\begin{theacknowledgments}
It is a pleasure to thank Joan Centrella for organizing this very interesting workshop,
Bill Keel for providing the optical image of NGC\,6240,
and G\"unther Hasinger and David Merritt for pleasant discussions.
\end{theacknowledgments}

\end{document}